\documentclass[%
 reprint,
superscriptaddress,
 amsmath,amssymb,
 aps,
 showkeys,
]{revtex4-1}

\usepackage[english]{babel}
\usepackage{graphicx}
\usepackage{dcolumn}
\usepackage{bm}
\usepackage{xcolor}

\begin{document}

\title{Enhancing the Refractive Index of Polymers with a Plant-Based Pigment}

\author{Mohammad Yasir}
  \thanks{These authors contributed equally, order determined by coin flip.}
  \affiliation{Department of Materials, ETH Z\"{u}rich, 8093 Z\"{u}rich, Switzerland.}
  
  \author{Tianqi Sai}
  \thanks{These authors contributed equally, order determined by coin flip.}
  \affiliation{Department of Materials, ETH Z\"{u}rich, 8093 Z\"{u}rich, Switzerland.}

\author{Alba Sicher}
  \affiliation{Department of Materials, ETH Z\"{u}rich, 8093 Z\"{u}rich, Switzerland.}
 \author{Frank Scheffold}
 \affiliation{Department of Physics, University of Fribourg, 1700 Fribourg, Switzerland }
 \author{Ullrich Steiner}
 \affiliation{Adolphe Merkle Institute, University of Fribourg, 1700 Fribourg, Switzerland }
 \author{Bodo D. Wilts}
 \affiliation{Adolphe Merkle Institute, University of Fribourg, 1700 Fribourg, Switzerland }
\author{Eric R. Dufresne}
  \email{eric.dufresne@mat.ethz.ch}
  \affiliation{Department of Materials, ETH Z\"{u}rich, 8093 Z\"{u}rich, Switzerland.}


\date{\today}

\begin{abstract}

Polymeric materials are prized for their formability, low density, and  functional versatility. 
However, the refractive indices of common polymers fall in a relatively narrow range between 1.4 and 1.6. 
Here, we demonstrate that loading commercially-available polymers with large concentrations of a plant-based pigment can effectively enhance their refractive index.
For polystyrene loaded with 67w/w\% $\beta$-carotene, we achieve a peak value of 2.2 near the absorption edge at $531~\mathrm{nm}$, while maintaining values above 1.75 across longer wavelengths of the visible spectrum. 
Despite high pigment loadings, this blend maintains the thermoforming ability of polystyrene, and $\beta$-carotene remains molecularly dispersed.  
Similar results are demonstrated for the plant-derived polymer ethyl cellulose. 
Since the refractive index enhancement is intimately connected to the introduction of strong absorption, it is best suited to applications where light travels short distances through the material, such as reflectors and nanophotonic systems.
We experimentally demonstrate enhanced reflectance from films, as large as  seven-fold for ethyl cellulose at selected wavelengths.  
Theoretical calculations that highlight that this simple strategy  can significantly increase light scattering by nanoparticles and enhance the performance of Bragg reflectors. 

\end{abstract}

\maketitle

High refractive index ($n$) materials are attractive for many high performance optical systems, including nanophotonics and solar cells.
While many inorganic materials naturally have very high refractive indices \cite{palik1998handbook,caseri2000nanocomposites}, polymeric alternatives are attractive due to their low density and high processability.
To that end, considerable effort has been invested in the development of high refractive index polymers and/or polymer composites (see \emph{e.g}
 \cite{liu2009,badur2018,jiang2017,kleine2016,chau2007,lu2009,tao2011,yetisen2016,kim2015,sanders2010,higashihara2015,song2016}).

Methods to prepare high refractive index materials can be classified into three approaches. 
First, monomers are chemically modified to incorporate heavy atoms such as sulfur \cite{olshavsky1995,kim2020,fang2020}. This can be synthetically tedious and costly. A polymer synthesized from vapor phase radical polymerization of elemental sulfur and 1,4-butanediol divinyl ether appears to have the highest refractive index reported for such a specialty polymer, with a value of about 2.1 at the wavelength of 400 nm, decaying to 1.9 at 700 nm \cite{kim2020}.
Second, high refractive index particles are dispersed throughout the polymer matrix. 
These are typically nanoparticles (NPs) of metal oxides such as zirconium oxide or titanium dioxide. This approach is challenging because  the particles need to be small and well-dispersed to avoid light scattering, while  their high surface energies tend to drive  agglomeration \cite{song2016,bockstaller2003,lu2009}. For these composite materials, refractive indices as high as 2.5  have been reported for 60\% of PbS in gelatin \cite{zimmennann1993}. Polymers loaded by metallic nanoparticles can also display an elevated refractive index in the vicinity of plasmon resonances, and display enhanced reflectance \cite{gehr1996}. 
Third, small absorbing molecules can be dispersed in the polymer matrix. To date, this approach has focused on pigments with absorption in the UV, like phenanthrene, providing modest increases of the refractive index, $\Delta n$, of about 0.07 across the visible range \cite{hanemann2011,hanemann2014,bohm2004,magrini2019}.

Pigments and other absorbing systems naturally can feature high refractive indices.
This is due the fact that both absorption and the refractive index are  derived from the same linear response function, the dielectric constant. 
Basic constraints on causality require the refractive index (related to the real part of the response function) and the absorption coefficient (related to the imaginary part) to be connected through the Kramers-Kronig relationship \cite{lucarini2005}.
This is famously exploited in silicon photonics, where the  refractive index is about 3.5 beyond the band gap \cite{reed2010,foster2006,schinke2015}.

Living systems can achieve very high refractive indices using organic pigments.  
Melanin has a refractive index of about 1.8 and is deployed in nanostructures by many species to produce structural colors
\cite{hill2006,yoshioka2011,stavenga2015}.
Artificial melanin has been used to create structurally colored biomimetic nanostructures \cite{kawamura2016,xiao2017}.
Pterin pigments, which absorb strongly at short wavelengths, have  recently been identified as an essential component of highly reflective structures in animals \cite{wilts2017, palmer2020}.
For example, the wing scales \emph{Pierid} butterflies carry micron-sized ellipsoidal particles made primarily from pterin pigments.
These particles have refractive indices as high as 2.9 at the edge of the absorption band, and remain above 2.0 at longer wavelengths \cite{wilts2017}. 
In this region of the spectrum, the high refractive index of these nanostructures leads to very efficient reflection. 

Here, we increase the refractive index of the common polymers polystyrene (PS) and ethyl cellulose (EC) by loading them the natural organic pigment $\beta$-carotene (BC).
We achieve refractive indices greater than 2 near the absorption edge, while maintaining the processability of the host polymer. 
All of these raw materials are produced commercially and made from naturally abundant elements (carbon, hydrogen and oxygen).
Moreover, BC and EC are plant-based edible materials. Thus, our high-refractive-index materials are cost-effective, sustainable, and environmentally friendly. 
Because of residual adsorption, these materials are best suited for applications where light travels no more than short distances through the material. 
We demonstrate the applicability of this approach with a combination of experiments and theory.
First, we measure increases of reflectance from flat interfaces as large as seven-fold greater than the base polymer. 
Second, we theoretically investigate the performance of nanoparticles and  nanostructures based on these materials.
Specifically, we find that pigment loading greatly enhances the reflectance of a multilayer structure and the scattering efficiency of nanoparticles.

\begin{figure}
  \includegraphics[width=8.5cm]{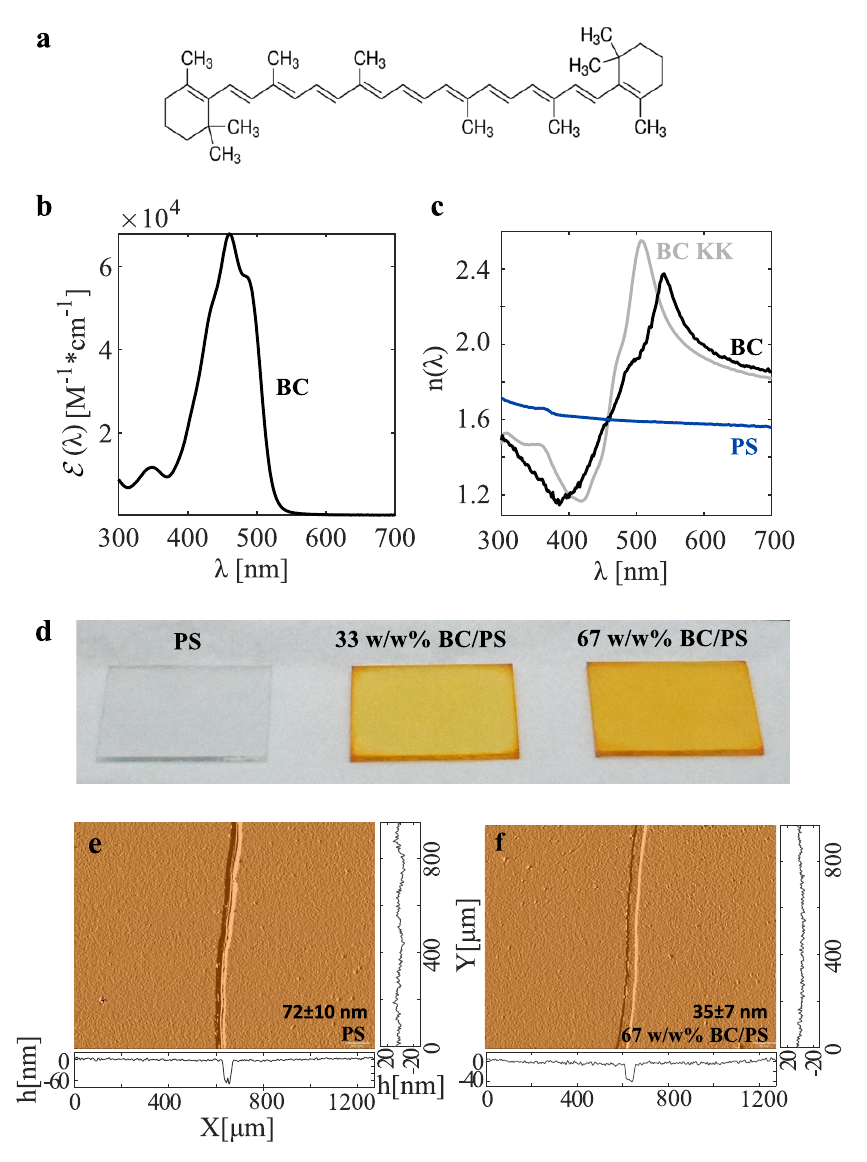}
  \caption{\label{fig:idea} \emph{Design principle.} (a) Chemical structure of BC. (b) Molar extinction coefficient of $0.1\,\mathrm{mM}$ BC in chloroform. (c) Measured refractive indices of the spin-coated pure BC film (black) and the spin-coated PS film (blue). Estimated refractive indices of pure BC by using the Kramers-Kronig (KK) relationship (gray). (d) Photos of spin-coated PS on glass slides with BC-loadings of 0w/w\%, 33w/w\% BC/PS and 67w/w\% BC/PS  (size: $2.5\times2.2~\mathrm{cm}^2$). (e,f) Profilometer images of PS, 33w/w\% BC/PS and 67w/w\% BC/PS films and corresponding height profiles.}
\end{figure}

In order to enhance the refractive index of polymeric materials using a pigment, it must satisfy two minimal criteria: (i) it must have a strong optical response and (ii) it can be dispersed in the desired polymeric matrix.

Since pigments are usually characterized by their absorption spectrum and not their refractive index, we applied the Kramers-Kronig relations (see \emph{e.g.} \cite{lucarini2005,sai2020}) to estimate the impact of dye loading on the refractive index. 
The change in the refractive index due to absorption can be estimated as:
\begin{equation}
    \Delta n(\lambda)=\frac{c}{2\pi^2} \int_{0}^{+\infty} \frac{\mathcal{E}(\lambda')}{1-\left(\lambda'/\lambda\right)^2}d\lambda'
   \label{kk}
\end{equation}
where $\mathcal{E}(\lambda)$ is the molar extinction coefficient of the pigment and \textit{c} is its molar concentration \cite{sai2020}. 
The numerical approach to the evaluation of this integral is given in the Supporting Information.

Qualitatively, the refractive index increases  with the molar extinction coefficient. Absorption over a narrow band of wavelengths results in a peak of the refractive index at the long-wavelength edge of the absorption band \cite{sai2020}. 
The refractive index decays slowly at longer wavelengths. 
An ideal pigment to enhance the refractive index would feature an absorption spectrum with a large integral and a sharp drop to zero at long wavelengths.

We identified $\beta$-carotene, shown in Fig. \ref{fig:idea}a, as a suitable candidate based on its absorption spectrum, shown in Fig. \ref{fig:idea}b. 
Solutions of BC in chloroform absorb strongly between $400~\mathrm{nm}$ and $550~\mathrm{nm}$, with a peak molar extinction coefficient of about $68000~\mathrm{M^{-1} cm^{-1}}$ at $460~\mathrm{nm}$.
In addition to these suitable absorption properties, BC is an healthy and abundant natural product, soluble in a variety of common solvents.

The absorption spectrum of BC implies that it should have a large refractive index.
To measure this, we spin-coated a $145~\mathrm{nm}$ thick film of BC on a silicon wafer and measured its refractive index using an ellipsometer. The measured refractive index is shown as the black line in Fig.~\ref{fig:idea}c.
A peak refractive index of $2.38$ is observed at a wavelength of $541\,\mathrm{nm}$. 
The refractive index stays above $2.0$ up to a wavelength of $600~\mathrm{nm}$,  well outside of its absorption band. 
At this wavelength, the residual absorption is $276\,\mathrm{M^{-1}cm^{-1}}$, about $250\times$ smaller than its peak value.
For comparison, the refractive index predicted by the measured absorption spectrum using Eq.~\ref{kk} is shown as a gray line in Fig.~\ref{fig:idea}c.
Note that the measured peak in the refractive index is weaker and at a longer wavelength than expected. We suspect that these differences arise from changes in the local electronic environment in solution and the solid state \cite{gong2018}.

While pure BC has attractive optical properties, it is difficult to process.
We hypothesized that dispersing BC in a polymer matrix could maintain a high index of refraction while improving its processibility.  
We selected polysytrene as a candidate polymer matrix. PS is a common thermoplastic polymer with a refractive index of about 1.6 across the visible spectrum \cite{sultanova2009}, as shown in Fig.~\ref{fig:idea}c.

\begin{figure*}
\centering
  \includegraphics[width=15cm]{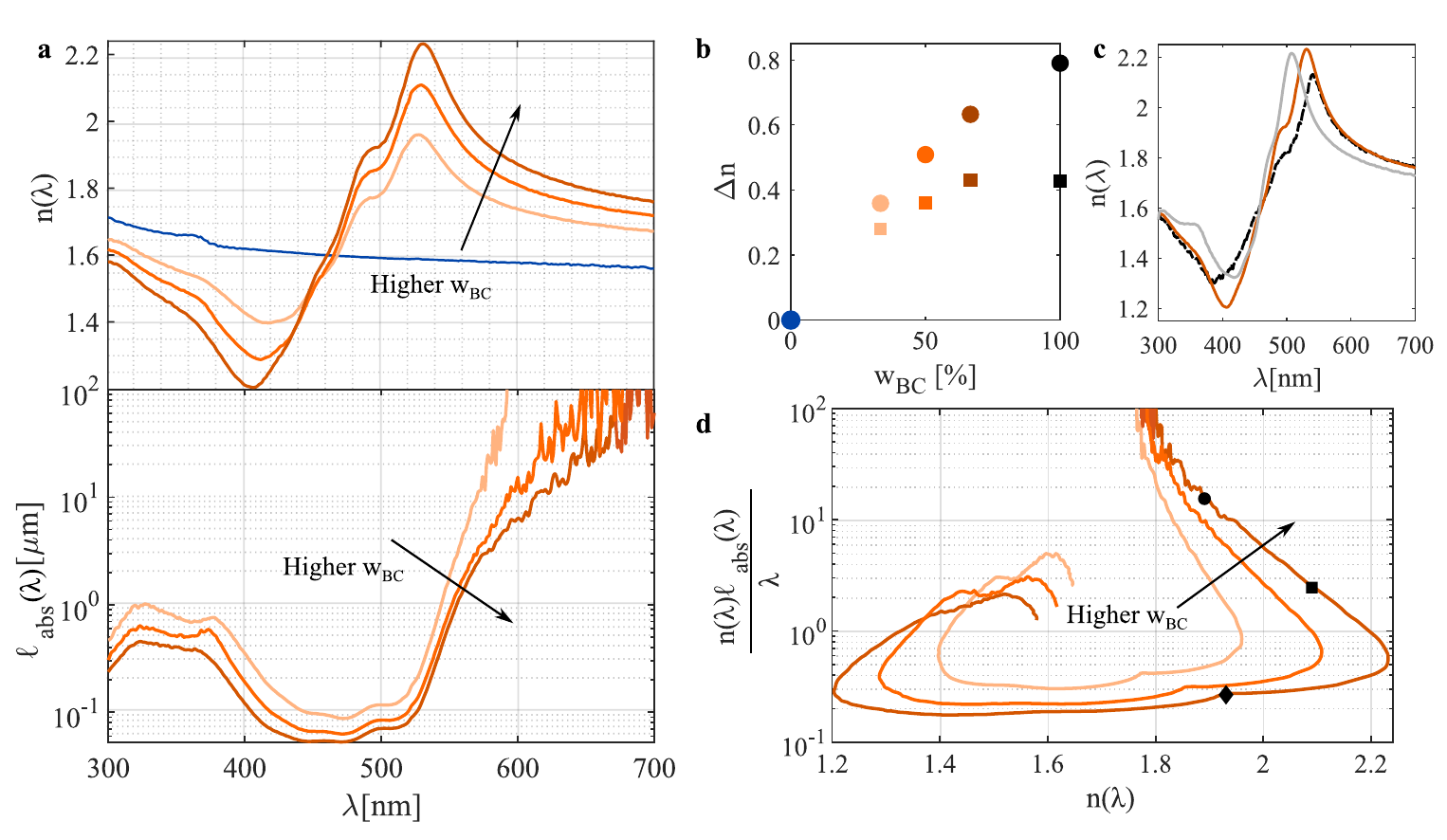}
  \caption{\emph{Trade-off between high refractive index and absorption.} (a) Measured refractive index dispersion and absorption lengths of BC-containing PS films with loading of 67w/w\%, 50w/w\% and 33w/w\% and pure PS (blue). (b) Peak refractive indices (circles) and refractive indices when $n(\lambda)l_{abs}=5\lambda $ (squares) for different weight percentage BC-containing PS films. (c) Calculated and measured refractive indices of 67w/w\% BC/PS film. Measured in solid, orange line, calculated by mixing rule in black, dashed line and calculated by Kramers-Kronig in light gray, solid line. (d) Normalized absorption length vs. refractive indices for 67w/w\% BC/PS film, 50w/w\% BC/PS film and 33w/w\% BC/PS film. Data points of 67w/w\% BC/PS for wavelengths of $500~\mathrm{nm}$, $535~\mathrm{nm}$, and $592~\mathrm{nm}$ are marked in diamond, square, and circle, respectively.}
  \label{fig:n_absl}
\end{figure*}

We prepared thin composite films of BC and PS by spin coating. Photographs of the samples are shown in Fig.~\ref{fig:idea}d.
We quantified the uniformity and thickness of the films using optical profilometry, as shown in Fig.~\ref{fig:idea}ef..
We saw no significant difference in the roughness of the films, even at very high loadings of BC.
The optical properties of a range of BC/PS films were characterized using ellipsometry.
The measured  refractive indices $n(\lambda)$ and absorption lengths $\ell_\mathrm{abs}(\lambda)$ are plotted in Fig. \ref{fig:n_absl}a.
As the weight percentage of BC increases, the refractive index dispersion shows strong anomalous dispersion, \emph{i.e.} a pronounced dip/peak develops on either side of the absorption band.
The highest, measured refractive index is $2.23$ at a wavelength of $531~\mathrm{nm}$ for a 67w/w\% BC/PS film.

The peak refractive index increases with the concentration of BC, as plotted in Fig. \ref{fig:n_absl}b. At concentrations lower than 67w/w\%, the concentration dependence is roughly linear, as expected by the Eq.~\ref{kk}.
The Kramers-Kronig relation also predicts the wavelength dependence of the refractive index of the 67w/w\% BC/PS blend rather well, as shown by the light gray curve in Fig. \ref{fig:n_absl}c.

The Kramers-Kronig calculation for the blend's optical properties is based on an extrapolation from the dye behavior in a dilute solution.
Alternatively, we can use the law of mixtures \cite{heller1965}, to extrapolate the blend's optical properties based on the measured refractive index of pure BC and PS.
In this way,
\begin{equation}
    n(\lambda) = n_{BC}(\lambda)\phi + n_{PS}(\lambda)(1-\phi)
    \label{eq_mix}
\end{equation}
where $\phi$ is the volume fraction of BC and $n_{BC}$, $n_{PS}$ are the refractive indices of pure BC and pure PS, respectively. 
Since  BC and PS have similar density, their weight and volume fractions are essentially the same. This estimate for the refractive index is shown as a dashed curve in Fig.~\ref{fig:n_absl}c,  and also matches the data reasonably well.

The possible applications of this approach for increasing the refractive index are limited by residual absorption.
The intrinsic trade-off between high refractive index and absorption is captured nicely by Fig.~\ref{fig:n_absl}d, where we plot the absorption length (in units of wavelength) versus the refractive index.
While the 67w/w\% BC/PS sample can reach refractive indices as high as $2.23$, the absorption length at the same frequency is only about half a wavelength. 
This rules out any application where light propagates through the material, such as lenses.
Luckily, with only a 10\% decrease of the refractive index, to a value of 2.0, the absorption length increases ten fold to about 5 wavelengths in the material.
While this may be too short for most applications, we will show below that it can be effectively exploited for nanophotonic applications.

\begin{figure*}
  \includegraphics[width=15cm]{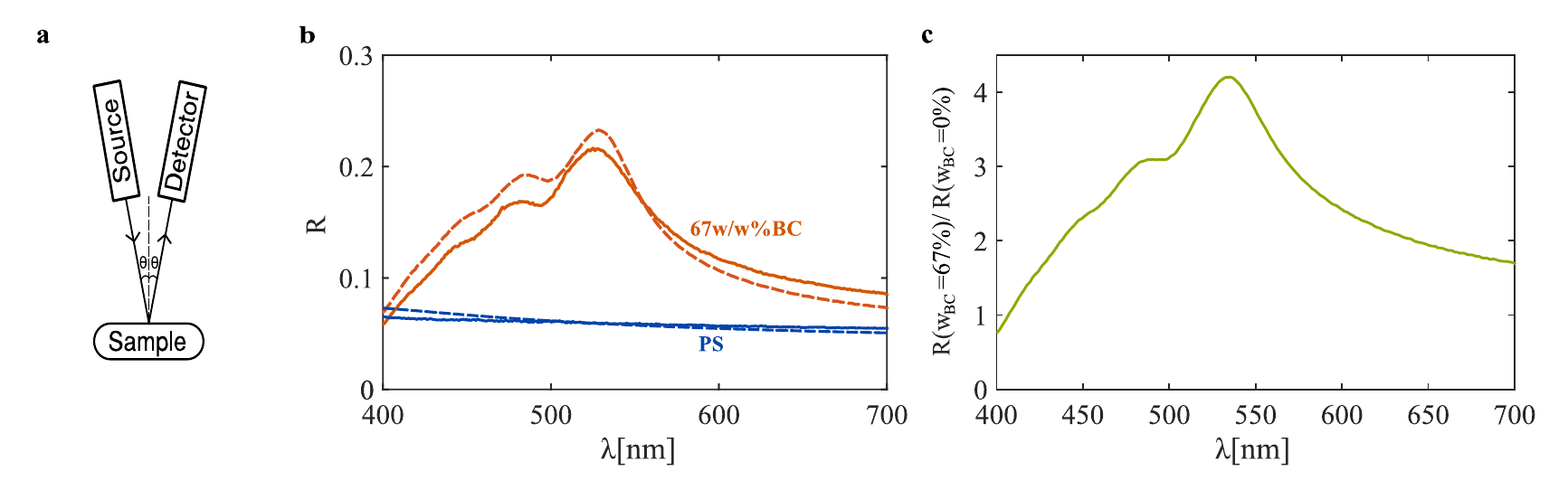}
  \caption{\emph{Reflectance increases significantly by incorporating BC.} (a) Schematic plot of the reflectance measurement set-up. ($\theta=15^\circ$) (b) Measured (solid lines) and modelled (dashed lines) reflectances of the 67w/w\% BC/PS film (orange) and the pure PS film (dark blue). (c) Theoretical reflectances ratio of the 67w/w\% BC/PS film ($57~\mathrm{nm}$) over the pure PS film ($57~\mathrm{nm}$).}
  \label{fig:R}
\end{figure*}

Counterintuitively, adding an absorbing material to a thin film increases its reflectance.
To demonstrate this, we compare the specular reflectance of PS and 67w/w\% BC/PS thin films in Fig.~\ref{fig:R}.
While the reflectance of the pure PS film is around $0.06$ across the visible spectrum, the reflectance spectrum of the 67w/w\% BC/PS film reaches a peak value of $0.22$ (Fig.~\ref{fig:R}b), about $4\times$ higher than the reflectance of a PS film of the same thickness (Fig.~\ref{fig:R}c). 
We fitted the measured reflectance spectra to Transfer-matrix calculations \cite{pascoe2001} using the measured refractive indices of both films, allowing their thickness to vary.
The resulting thicknesses are $57~\mathrm{nm}$ and  $41~\mathrm{nm}$ for PS and BC/PS films, respectively. The best fit spectra fit the data reasonably well.
For comparison, the optical profilometer thickness measurements of the same films thickness were $72\pm10~\mathrm{nm}$ and $35\pm7~\mathrm{nm}$ for PS and BC/PS, respectively.

\begin{figure}
  \includegraphics[width=8.5cm]{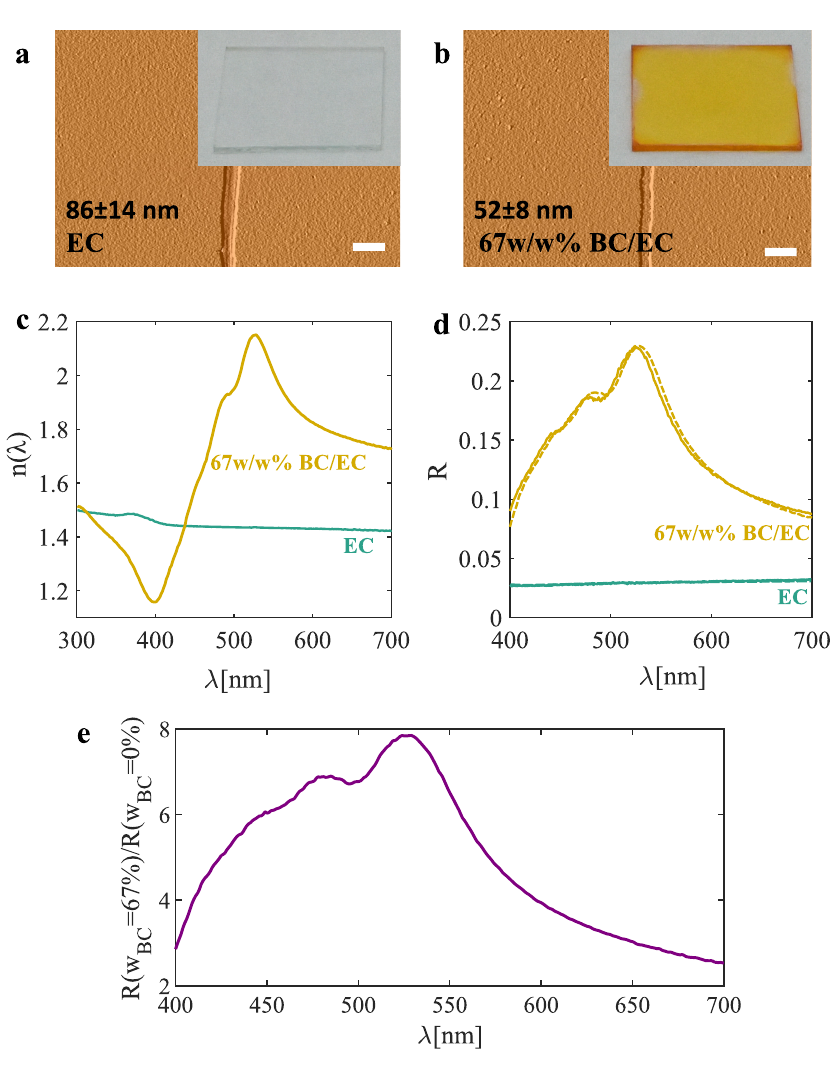}
  \caption{\emph{High refractive index blend of plant-based pigment, BC and plant-based polymer, EC.} (a,b) Photos and surface profiles of EC and 67w/w\% BC/EC films. (c) Measured refractive indices of 67w/w\% BC/EC film (orange), and pure EC (green). (d) reflectance of the 67w/w\% BC/EC film (orange) and the pure EC film (green). Measured results are in solid lines. Modelled results are in dashed lines using thicknesses of $58~\mathrm{nm}$ and $45~\mathrm{nm}$ for BC/EC and EC respectively. (e) Ratio of measured reflectance of the EC/BC film to the theoretical reflectance of a EC film with the same thickness ($58~\mathrm{nm}$).} 
  \label{fig:Et}
\end{figure}

To demonstrate the generality of our approach, we produced thin films of ethyl cellulose EC loaded with 67w/w\% BC.
EC is an attractive choice, because it is an inexpensive, commercially available, abundant plant-sourced polymer. We found very similar results to the PS films shown above, as summarized in Fig.~\ref{fig:Et}.
Thin films of EC and BC/EC had similar roughness and uniformity (see Fig.~\ref{fig:Et}ab). 
While pure EC has a refractive index of about 1.44 across the visible wavelength range, the refractive index of the BC/EC composite had a peak refractive index of 2.15 at $528~\mathrm{nm}$, decaying to 1.83 at $600~\mathrm{nm}$, as shown in Fig.~\ref{fig:Et}c.
The specular reflection from thin films of BC/EC was almost eight-fold higher than plain EC films at the peak near $529~\mathrm{nm}$ (Fig.~\ref{fig:Et}de).
It should be noted that these high refractive index films are based solely on plant-based materials.
 
\begin{figure*}
  \includegraphics[width=15cm]{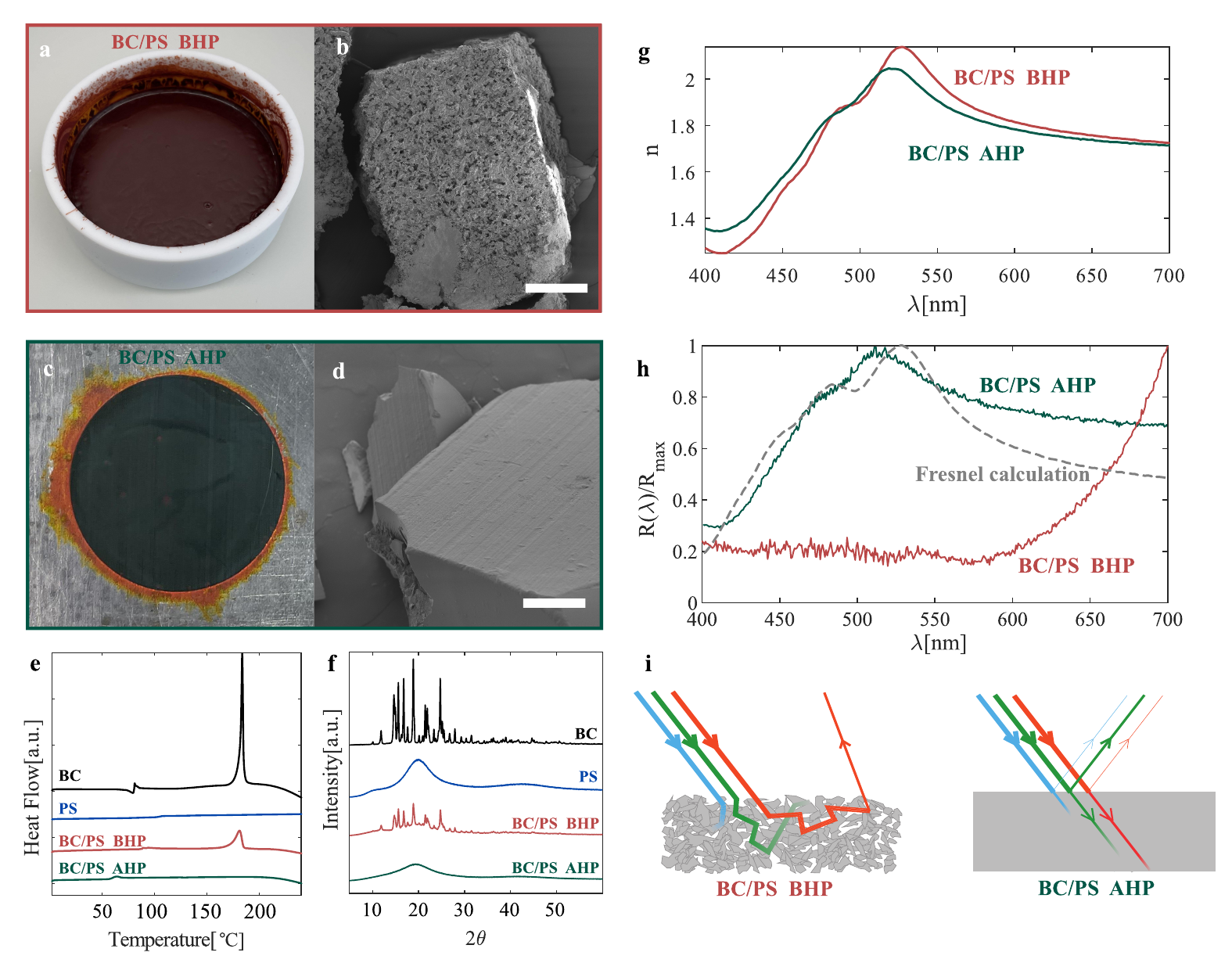}
  \caption{\emph{Processability, structure, and optical properties of BC/PS composites}. (a) Photo of 40w/w\% BC/PS mixture after solution casting, and before hot pressing (BHP). Diameter of the petri dish is $10~\mathrm{cm}$. (b) SEM image of the sample. Scale bar: $200~\mathrm{\mu m}$. (c) Photo of 40w/w\% BC/PS composite after hot pressing (AHP) at $170^{\circ}C$. Diameter of the green disc is $2.5~\mathrm{cm}$. (d) SEM image of the sample after hot pressing at $170^{\circ}C$. Scale bar: $200~\mathrm{\mu m}$. (e),(f) DSC and XRD results of a 40w/w\% BC/PS mixture before (BHP) and after (AHP) hot pressing. (g) Refractive index dispersions of 40w/w\% BC/PS composites before (red) and after hot pressing (green). (h) Normalized reflectances of the 40w/w\% BC/PS mixture before (red) and after (green) hot pressing. reflectance calculated by the Fresnel equations in gray, dashed line. (i) Schematic of the light propagation in BC/PS films.}
  \label{fig:process}
\end{figure*}

The results above demonstrated that BC can be incorporated at high-loadings in a polymer matrix to achieve a strong optical response. However, it remains to be seen whether these mixtures maintain the processability of the host polymer.
We therefore tested the thermoforming ability of 40w/w\% BC/PS mixtures. For this, we made a 40w/w\% BC/PS mixture by solution casting using dichloromethane as solvent under vacuum ($180\,\mathrm{mbar}$) at room temperature.
The resulting composite is dark orange in color (Fig.~\ref{fig:process}a), and has a porous structure as revealed by SEM (Fig.~\ref{fig:process}b).
Then, we hot pressed the composite at $170^{\circ}\mathrm{C}$, which resulted in the formation of a $2\,\mathrm{mm}$ thick film.
The resulting material was dark green (Fig.~\ref{fig:process}c), and no porosity was visible in SEM images (Fig.~\ref{fig:process}d). 

Differential scanning calorimetry (DSC, Fig.~\ref{fig:process}e) and x-ray diffraction (XRD, Fig.~\ref{fig:process}f) revealed that the BC was in a crystalline form after solution casting, but not after hot pressing.
Specifically, DSC revealed a melting event in the solution-cast BC/PS sample (before hot pressing (BHP)) between $170-190^{\circ}C$, matching the exothermic peak in the pure BC sample.
Similarly, XRD revealed a number of sharp, crystalline peaks in the solution-cast BC/PS sample that matched the diffraction peaks of pure BC.
These characteristic features of crystalline BC are completely absent from the DSC and XRD results after hot pressing (AHP).
We therefore conclude that hot-pressing the BC/PS film at the melting point of BC dissolves it into the PS matrix.
Upon cooling, the PS vitrifies before the BC re-crystallizes, leaving a solid BC solution frozen in vitrified PS. 
This is particularly attractive for optical applications because molecularly dispersed pigments do not scatter light.
To confirm that heat treatment did not compromise optical performance, we measure the refractive index dispersion of the hot pressed material, shown in Fig.~\ref{fig:process}g, 
by ellipsometry on a thin film formed by spin coating after redissolving the hot-pressed sample.
We observed only a slight decrease in the peak refractive index, with a maximum value of $2.04$ at $520~\mathrm{nm}$. Processing at higher temperatures ($230^{\circ}C$) enabled extrusion of the material into fibers, but almost entirely removed the refractive index enhancement, as shown in Fig.~\ref{fig:procsi}, likely due to a chemical decomposition of the BC.

Hot-pressing at $170^{\circ}\mathrm{C}$ transforms the dark-red solution-cast mixture in to a dark green material.
To quantify this difference in appearance, we measured the specular reflectance of each sample, shown in Fig. \ref{fig:process}h.
While the solution-cast film only reflects the longest visible wavelengths, the reflectance of the hot-pressed film closely follows the profile of its measured refractive index, which both feature a peak near $520~\mathrm{nm}$.
For comparison, we superimpose the Fresnel prediction for specular back-reflection as a gray curve, where $R= |(\tilde{n}-1)/(\tilde{n}+2)|^2$, where $\tilde{n} = n + ik$ is the complex refractive index.

The appearance of the hot-pressed film is determined by its refractive index, while the color of the solution cast film is dominated by its absorption spectrum.
As shown above, the solution cast film has a complex structure, which apparently has three phases, a PS-rich phase, BC crystals, and air.
This complex structure should strongly scatter light of all wavelengths, characterized by the transport mean-free path, $\ell_\mathrm{t}$.
When an absorption-free sample is much thicker than $\ell_\mathrm{t}$, it appears white because each wavelength can be scattered multiple times and re-emerge from the surface. 
In the presence of absorption, only wavelengths with an absorption length, $\ell_\mathrm{abs}$, much longer than $\ell_\mathrm{t}$ will be reflected, as shown schematically in Fig.~\ref{fig:process}i.
In our BC/PS samples, the absorption length increases steadily at long-wavelengths (see Fig.~\ref{fig:n_absl}a).
With a presumed scattering length on the order the wavelength of light, only the longest wavelengths can be scattered enough times to be reflected from the sample. 
Therefore, the color of the strongly-scattering solvent-cast sample is dominated by the absorption properties of the medium.
This is the same color-producing mechanism as pigmentary particles. 
On the other hand, hot-pressing transforms the mixture in a single phase as confirmed by SEM, DSC, and XRD measurements of Fig.~\ref{fig:process}.
With no internal scattering structures, the reflection is limited to the first surface, and the rest of the light propagates through the material without scattering until it is eventually absorbed, as shown schematically in Fig. \ref{fig:process}i.

\begin{figure*}
  \includegraphics[width=12.5cm]{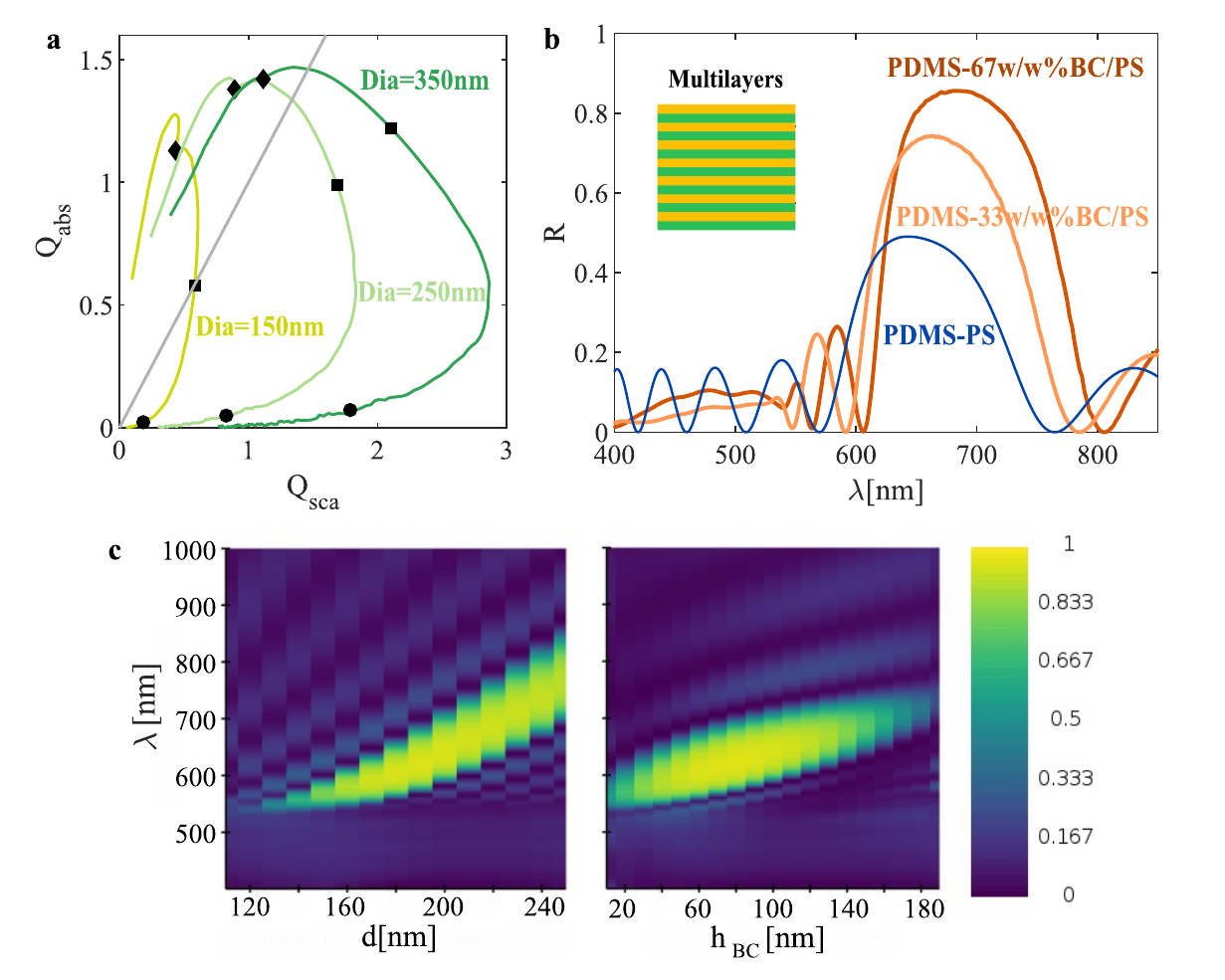}
  \caption{\emph{Numerical simulations of properties of nanophotonic structures made from BC/PS composites}. (a) Scattering efficiencies versus absorption efficiencies for 67w/w\% BC/PS particles with different diameters (Dia) in water, calculated by Mie theory. Data points for wavelengths of $500~\mathrm{nm}$, $535~\mathrm{nm}$, and $592~\mathrm{nm}$ are marked in diamonds, squares, and circles, respectively. (b) Modelled reflectances of a multilayer Bragg reflector with seven repeated layers of BC-loaded PS (thickness of each layer $100~\mathrm{nm}$) and a material with $n=1.4$ (thickness of each layer $120~\mathrm{nm}$). (c) reflectances of 7-repeat multilayers of 67w/w\% BC/PS and a material with $n=1.4$, where the thicknesses of the layers are varied. In the left panel, the thickness of the BC/PS layer, $h_\mathrm{BC}$ is fixed at $100~\mathrm{nm}$ and total thickness of each repeat unit, $d$, is varied. In the right panel, $d$ is fixed at $200~\mathrm{nm}$ and $h_\mathrm{BC}$ is varied. }
  \label{fig:theory}
\end{figure*}

So far, we have demonstrated the efficacy of enhancing the refractive index by pigment loading for simple films, where we demonstrated a strong effect on their reflectance (Figs.~\ref{fig:R},\ref{fig:Et}, and \ref{fig:process}h).
To investigate whether dye-loaded polymers could be attractive for nanophotonic applications, we investigated the impact of dye-loading on the scattering of individual nanoparticles and the efficiency of multilayered Bragg reflectors. 

To predict the influence of the refractive index enhancement on the scattering of particles made of BC/PS mixtures,
we calculated the scattering and absorption properties of individual particles using Mie theory \cite{bohren2008}. 
Fig.~\ref{fig:theory}a compares the scattering efficiency, $Q_{sca}$, to the absorption efficiency $Q_{abs}$, for 67w/w\% BC/PS nanoparticles in water.
The scattering (absorption) efficiency $Q_{sca}$ ($Q_{abs}$) is the ratio of the scattering (absorption) cross section to the geometrical cross section. For $350~\mathrm{nm}$ diameter particles, the maximum scattering efficiency is $2.86$ at a wavelength of $552~\mathrm{nm}$.
While the absorbtion efficiency is much smaller under the same conditions, it is still significant ($Q_{abs}=0.56$). 
Note that $Q_{sca}$ is a factor of five larger than the scattering efficiency of a pure PS particle at the same size and wavelength (Fig. S3). 
In \emph{Pierid} butterflies, a similar enhancement has been observed. 
There, incorporation of pterin pigments increases the scattering efficiency of microparticles  ten-fold   \cite{wilts2017}. 
Further work is needed to reveal how this competition of scattering and absorption translates to the optical properties of periodic and quasi-periodic distributions of such particles in photonic crystals \cite{lopez2003,joannopoulos1997} and photonic glasses \cite{magkiriadou2014,schertel2019a}. 

Putting aside the complexity of single particle response, we can assess the impact of absorption-enhanced scattering in a one-dimensional periodic structure, the multilayer Bragg reflector (MBR), shown schematically in Fig.~\ref{fig:theory}b. We investigated the optical response of a multilayer with seven alternating layers of two materials. 
One layer has a refractive index of 1.4 (\emph{e.g.} PDMS), and the second has the refractive index dispersion of our BC/PS mixture.
Keeping the layer thicknesses fixed ($120~\mathrm{nm}$ for the low index material, and $100~\mathrm{nm}$ for BC/PS blends), we varied the loading of BC from 0 to 67\% and calculated the reflectance spectrum of the structure at normal incidence, shown in Fig.~\ref{fig:theory}b. 
Without further optimization, the peak reflectance increases from 0.48 with pure PS up to 0.85 with 67w/w\% BC/PS. 
Varying the thickness of the layers, the reflectances of this system can reach as high as 0.95, as shown in Fig. \ref{fig:theory}c.
We observe a cut-off effect for smaller lattice spacings.  
As the reflectance peak reaches the absorption band, it first narrows and then fully disappears.
Specifically,  the reflectance peak width is around $100~\mathrm{nm}$ for $d=180~\mathrm{nm}$, narrowing to $50~\mathrm{nm}$ for $d=160~\mathrm{nm}$. 

We have shown that BC can be dispersed in PS and EC at very high loadings. These mixtures can reach refractive indices over 2 for a range of wavelengths, and display enhanced reflectance in thin films. Composites of BC and PS maintain their thermoforming ability. Thermoforming has an additional benefit: it dissolves BC crystals, preventing unwanted scattering from crystallized pigment. 
Numerical calculations suggest that the enhancement of refractive index due to the addition of dye can significantly enhance scattering by nanoparticles and improve the reflectance of Bragg reflectors.

This work demonstrates a new degree of freedom for the design of photonic nanostructures, enabling the enhancement of refractive index at specific wavelengths. 
Our theoretical results on Bragg reflectors suggest this strategy is particularly effective when wavelengths with strongly-enhanced refractive indices overlap with the wavelengths selected by the nanostructure.   This work demands further development of the theory of light transport in nanostructures with simultaneous scattering and absorption. To realize the photonic nanostructures proposed here, several challenges in materials chemistry and processing must be overcome. For example, new synthetic approaches are required to produce monodisperse polymer colloids with high dye loadings. Additionally, our proof-of-concept pigment, $\beta$-carotene, is prone to oxidation, and strategies to stabilize the pigment would greatly improve the longevity of these materials.

\medskip
\textbf{Acknowledgements} \par 
We thank Kiril Feldman for assistance with polymer processing and DSC measurements, Thomas Weber for help with XRD, Hui Cao for feedback on the optical data and Carla Fernandez-Rico for comments on the manuscript.
This work was financially supported by the Swiss National Science Foundation through the NCCR Bio-inspired Materials. 

\section*{Materials and Methods}

\subsubsection*{Materials}
BC (synthetic, $\geq$93\% (UV)), PS (35 kDa (used for making films atop silicon wafers and glass slides) and 192 kDa (used for processability experiments)), EC (48.0-49.5\%~(w/w) ethoxyl basis), chloroform and dichloromethane were bought from Sigma-Aldrich and used as received.

\subsubsection*{UV-VIS spectrometry}
The absorption spectrum of chloroform and BC solution in chloroform (0.1 mM) was recorded by a UV-Vis spectrophotometer (Agilent Cary 60). The absorbance of chloroform was subtracted from the absorbance of BC solution in order to have absorbance of pure BC. This absorbance data was used to calculate the molar extinction coefficient of BC according to the Beer-Lambert law.

\subsubsection*{Film Preparation}
To a weighed amount of PS or EC, the BC was added according to w/w\% (Table 1). A solution of this mixture in chloroform at a given concentration (Table 1) was stirred at 800 rpm under dark at room temperature. This solution was filtered through a 0.45 µm filter and the film was spin coated at 1000 rpm for 10 min over silicon wafer (size: $2\times2~\mathrm{cm}^2$) or glass slide (size: $2.5\times2.2~\mathrm{cm}^2$). The volume of solution taken for spin coating was 250 or 300 µl for silicon wafer or glass slide, respectively.
Specific compositions of solutions used for spin coating are found in the Supporting Information.

\subsubsection*{Profilometry}
The imaging of the films using a 3D optical profilometer (S neox) using 10x DI lens. In order to know the thickness of the films, a cut was made in the film using blade to scrap the portion of the material. The difference between the heights of the planes with and without the material gave the thickness of the films.  

\subsubsection*{Ellipsometry}
For refractive index measurements, a spectroscopic ellipsometer (M2000, J.A. Woollam) was used to measure the phase and amplitude change of light after interacting with the samples.
All measurements were performed between 250 and $1000~\mathrm{nm}$ at an angle of incidence $65^{\circ}$, and all data were acquired and analyzed using WVASE software. Since the films are transparent at longer wavelengths, the films were regarded as a homogeneous material with the thickness fitted by Cauchy dispersion relation in the wavelength range from 600 to $1000~\mathrm{nm}$, After the film thickness was determined and fixed, the optical constants, refractive indices $n$ and extinction coefficient $\kappa$ were fitted using point to point fitting mode.
 
\subsubsection*{Reflectance measurement}
The specular reflectance of the films have been measured with an angle-resolved spectrophotometry setup. The light source is a deuterium-halogen lamp (Ocean Optics DH-2000-BAL), ranging from $200~\mathrm{nm}$ to $1000~\mathrm{nm}$. The sample is placed on a rotating stage and can be tilted to the desired angles. Finally, a light detector (Ocean Optics, QE Pro) is mounted on a rotating arm, which is controlled by software. The incident angle is set to be $15^\circ$ and the specular reflectance is measured. The reflectance is calculated using reflection from a broadband dielectric mirror Thorlabs, BBSQ2-E02) at the same incident angle as reference. During the reflectance measurement, a layer of translucent Scotch tape was applied on the back side of the glass substrate to suppress unwanted back reflection from the glass-air interface.

\subsubsection*{Optical modelling}
The measured reflectances of thin films were fitted by a transfer matrix method based on the RTACalc code developed by A. Rose. We assume the films are in air and on a semi-infinite glass substrate in the modelling.

The Mie scattering calculations were done using the open source MATLAB code developed by Christian Mätzler.

The reflectances of multilayer were simulated with 3D finite-difference time-domain (FDTD) calculations using Lumerical FDTD Solutions, a commercial-grade Maxwell equation solver.

\subsubsection*{XRD measurement}
Samples were finely ground into powders before the XRD measurements. The XRD measurements were performed on the Panalytical X'Pert PRO MPD diffractometer with Cu K$\alpha$ ($8.04~\mathrm{keV}$) radiation. 

\subsubsection*{DSC measurement}
The DSC measurements of the samples were permormed on TA DSC2500 with a ramping speed of $10^{\circ}\mathrm{C}$/min under inert atmosphere.

\subsubsection*{Scanning electron microscopy}
The scanning electron microscopy (SEM) observations were performed on the Hitachi SU5000.

\section{Supporting Information} \par 

\renewcommand{\thefigure}{S\arabic{figure}}
\renewcommand{\thetable}{S\arabic{figure}}

\setcounter{figure}{0}
\setcounter{table}{0}

\begin{table}[ht]
 \caption{Solutions for Spin Coating.}
 \centering
 \begin{tabular}{c}
 \includegraphics[width=8.5 cm, clip=true, trim=0cm 13.5cm 0cm 1.5cm]{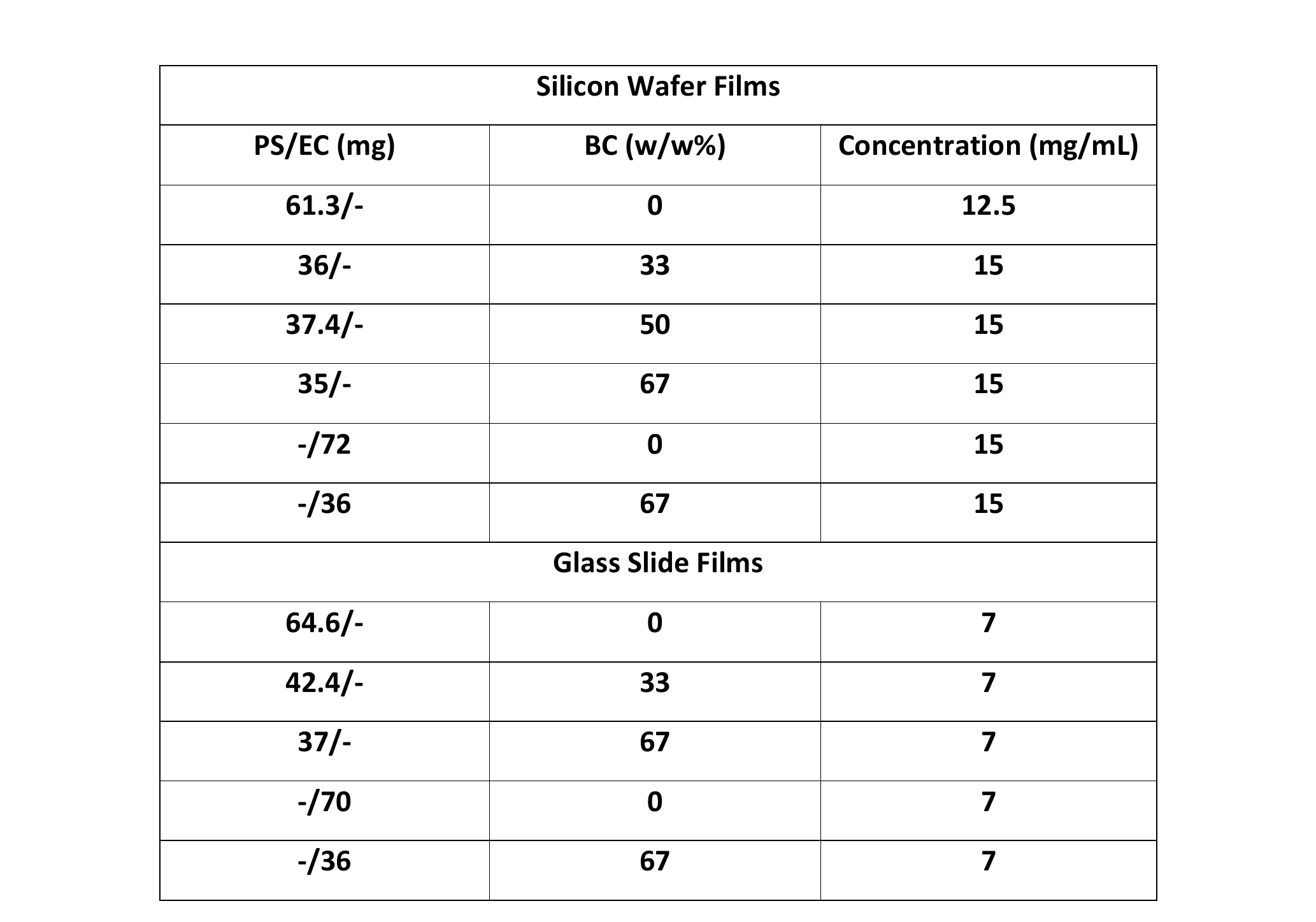}
 \end{tabular}
 \label{tab:}
\end{table}

\subsection{Refractive index prediction by Kramers-Kronig relations}

If we add a pigment to an optical system, according to the Kramers-Kronig relations, the resulting contribution of the pigment absorption to the refractive index $n(\lambda)$ can be precisely calculated if the absorption information for the entire spectral wavelength range ($0\to+\infty$) is known. Since we  measure only the   UV-Vis absorption spectrum, we determine   its contribution to the refractive index: 

\begin{equation}
    \Delta n'(\lambda)=\frac{c}{2\pi^2} \int_{200}^{800} \frac{\mathcal{E}(\lambda')}{1-(\lambda'/\lambda)^2}d\lambda'.
\end{equation}

The contribution of the missing spectral information below $200\,\mathrm{nm}$ and above $800\,\mathrm{nm}$ can be estimated by a Cauchy term $\Delta n_{m}$ as explained in our previous work\cite{sai2020}. 

In this work, we  mix large fractions of $\beta$-carotene with base polymer, with refractive index $n_b$. With the substitution of base polymer by $\beta$-carotene, the contribution of the substituted part of base materials on refractive indices need to be subtracted, which can be represented by $\Delta n_{b}$, which is negative. The base polymers, polystyrene and ethyl cellulose are transparent in the visible range. However, polystyrene and $\beta$-carotene share the same elements (C,H) and bondings (C=C,C-H,C-C). While the absorption in the UV-Vis range of organic molecules consisting of only C and H is mainly caused by the conjugation of C=C, the absorption outside UV-Vis results from transitions and vibrations of isolated bonds. So for a BC/PS system, $\Delta n_{b}$ and $\Delta n_{m}$ cancel out. Therefore, the overall refractive index of BC/PS composite $n_{c}$ can be written as:

\begin{equation}
    n(\lambda)= n_b(\lambda)+\Delta n'(\lambda)
\end{equation}

\begin{figure*}[h]
  \includegraphics[width=\linewidth]{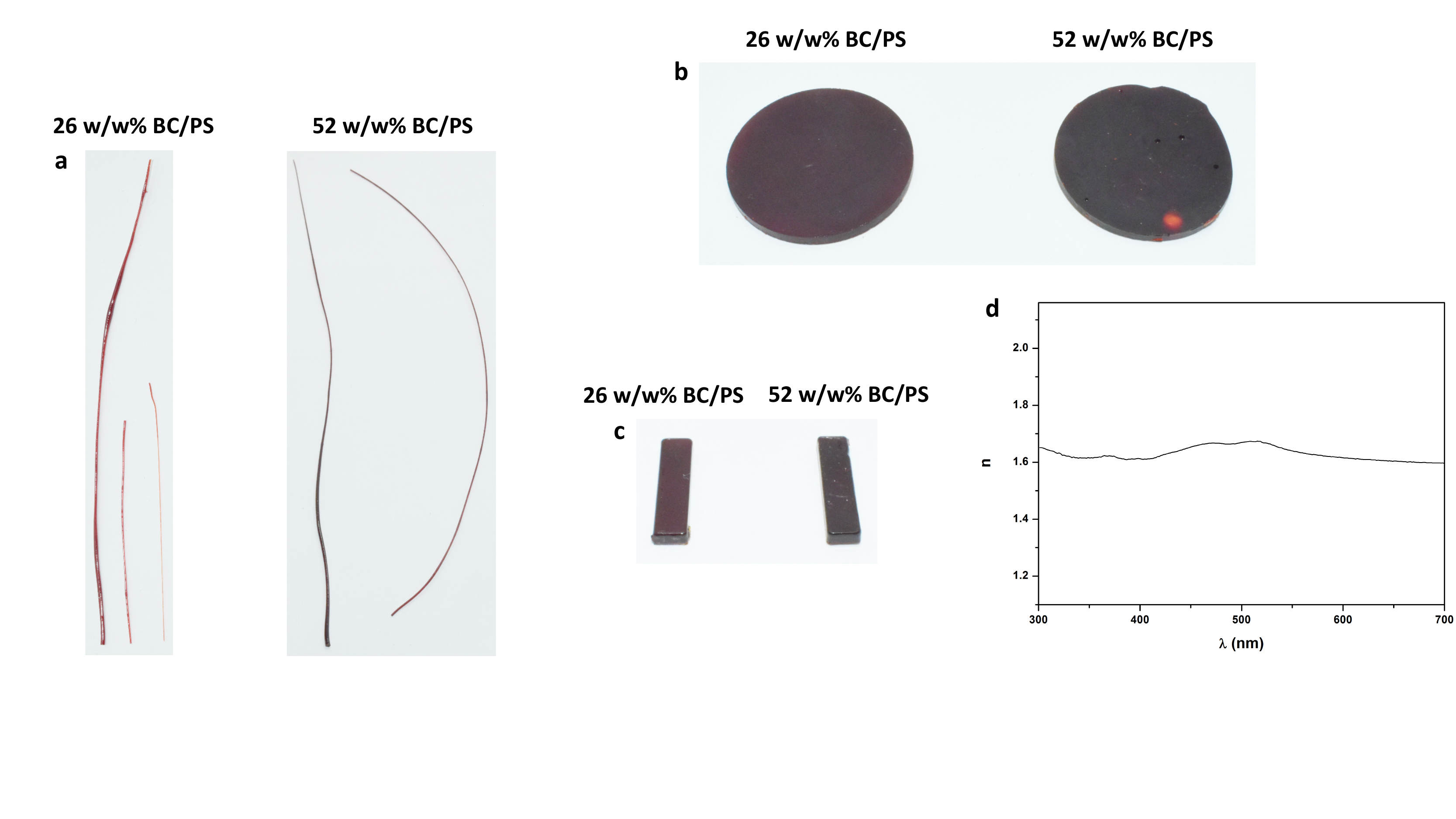}
  \caption{Processability of BC/PS composites by (a) extrusion, (b) hot pressing and (c) laser cutting. (d) Refractive index of 52w/w\% BC/PS composite after hot pressing at $205^{\circ}C$.}
  \label{fig:procsi}
\end{figure*}

\subsection{Processability}
A mixture of BC and PS was extruded into fibers having 26w/w\% and 52w/w\% BC/PS at $230^{\circ}C$ (Fig. S1a, Supporting Information). These fibers were hot pressed to form 2 mm thick circular films having the diameter of 30 mm at $205^{\circ}C$ (Fig. S1b, Supporting Information). These films were laser cut to give rectangular films of the same thickness having 20 mm length and 5 mm width (Fig. S1c, Supporting Information). Although the refractive index of the 52w/w\% BC/PS film formed by spin coating after redissolving the hot-pressed sample was decreased to 1.67 (maximum value at the wavelength of $510\,nm$) (Fig. S1d) due to the decomposition of most of the BC (because of heating at $230^{\circ}C$ during extrusion, a known phenomenon) \cite{valerio2021}. But these experiments show that the composites are processable at such high loadings of BC. As the amounts of BC and PS fed into the extrusion machine were unknown, $^1$H NMR spectra were recorded and used to determine the amount of BC and PS in the extruded fibres (Fig. S2).
\begin{figure*}[h]
  \includegraphics[width=\linewidth]{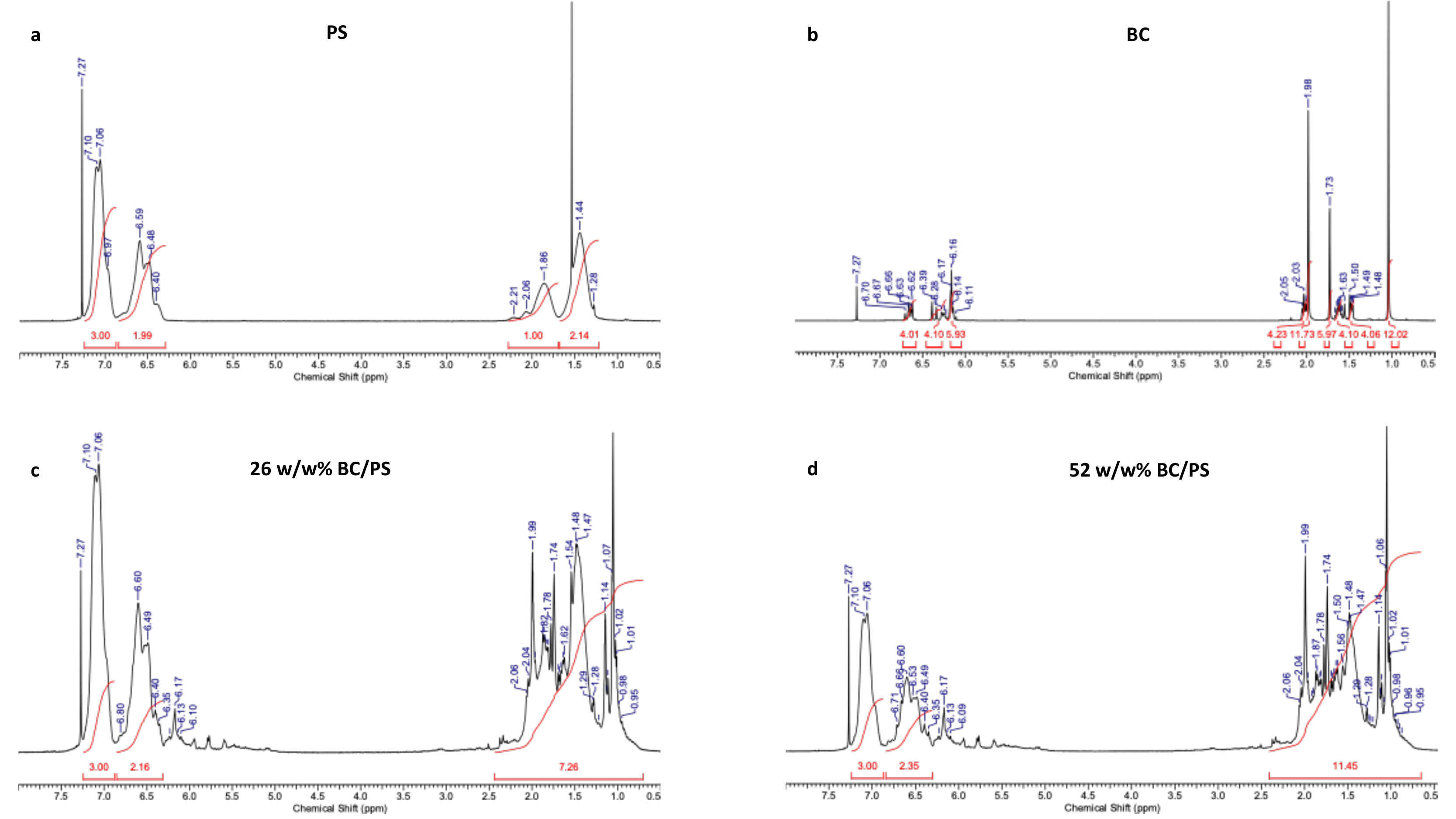}
  \caption{$^1$H NMR spectra (300 MHz, chloroform-d) of (a) PS, (b) BC, (c) {26w/w\%} BC/PS composite and (d) 52w/w\% BC/PS composite.}
  \label{fig:nmrspectra}
\end{figure*}

\subsection{Mie scattering efficiency enhancement}
\begin{figure*}[h]
  \includegraphics[width=0.5\linewidth]{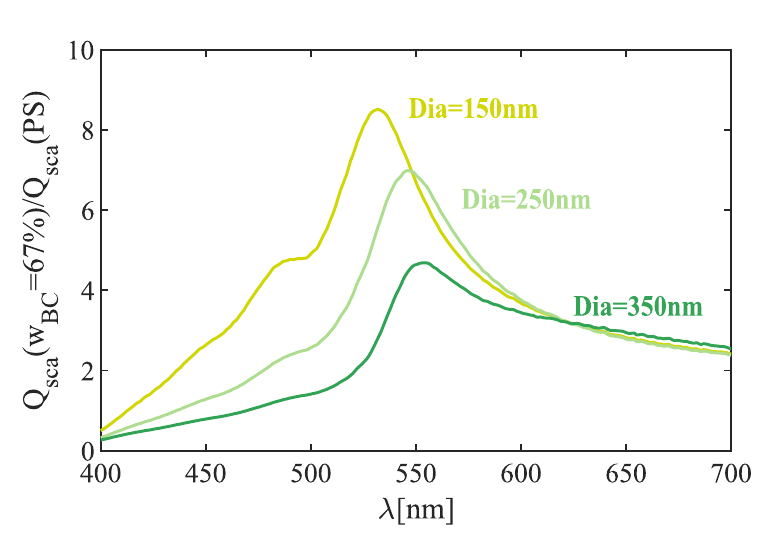}
  \caption{Enhancement factor of the Mie scattering efficiency $Q_{sca}$ of particles with the refractive index of 67w/w\% BC/PS relative to particles with the refractive index of PS..}
  \label{fig:q}
\end{figure*}

Scattering of particles can be strongly enhanced, depending on size and refractive index according to the calculations using Mie theory. For particles consisting of polystyrene, the scattering efficiency has a maximum at short wavelengths. For particles with the refractive indices of 67w/w\% BC/PS, the scattering of visible light is strongly enhanced, with a pronounced maximum in the visible wavelength range. This is particularly pronounced outside the absorption wavelength range of BC and leads to dramatic increases in the scattering efficiency of up to a factor of 8 (Fig.\ref{fig:q}). 

\medskip
%

\end{document}